\documentstyle[sprocl,epsfig,epsf]{article}

\def\be{\begin{equation}}
\def\ee{\end{equation}}
\def\bea{\begin{eqnarray}}
\def\eea{\end{eqnarray}}

\def\beq{\begin{equation}}
\def\eeq{\end{equation}}
\def\bea{\begin{eqnarray}}
\def\beaa{\begin{eqnarray*}}
\def\eea{\end{eqnarray}}
\def\eeaa{\end{eqnarray*}}
\def\bq{\begin{quote}}
\def\eq{\end{quote}}
\def\gappeq{\mathrel{\rlap {\raise.5ex\hbox{$>$}} {\lower.5ex\hbox{$\sim$}}}}

\def\lappeq{\mathrel{\rlap{\raise.5ex\hbox{$<$}} {\lower.5ex\hbox{$\sim$}}}}

\begin{document}

\title{Theoretical Introduction to Physics with Linear Colliders\\}

\author{ John Ellis }

\address{Theory Division, Physics Department,
 CERN, 1211 Geneva 23, Switzerland}


\maketitle\abstracts{
The major open questions in particle physics are summarized, as are the 
abilities of linear colliders of different energies to add to the knowledge 
obtainable from the LHC in various scenarios for physics beyond the 
Standard Model. A TeV linear collider would provide much additional 
insight into electroweak symmetry breaking, in particular, and a multi-TeV 
linear collider would add even more value in all the scenarios studied, 
for example in models with supersymmetry or extra dimensions.\\
~~\\
\begin{center}
{CERN-PH-TH/2004-166}\\
~~\\
{\it Invited Talk presented at the International Conference on Linear 
Colliders,\\
Paris, April 19-23, 2004}
\end{center}
}

\section{Open Questions beyond the Standard Model}

The primary justification for any new accelerator must be its capability
to explore and understand new physics beyond the Standard Model.
Motivating the searches for such new physics, there is a long list of
fundamental questions raised but left unanswered by the Standard Model,
including: What is the origin of particle masses? Are they due to a Higgs
boson, and is this accompanied by other new physics? Why are there so many
different types of matter particles? Are the fundamental forces unified?
What is the true quantum theory of gravity? There are plenty of topics
where LCs of different energies can contribute!

\section{The Physics Case for a LC}

The LHC will make the first exploration of the TeV energy range, and is
confidently expected to discover the Higgs boson, if it 
exists~\cite{LHCH}. It is also
likely to provide some evidence of whatever replaces it, if the Higgs
boson does not exist. The LHC has also been shown to have excellent
capabilities to search for other new physics that might accompany a Higgs
boson around the TeV scale, such as supersymmetry and/or extra dimensions.

Many studies have also shown that a LC can add significant value to this
initial exploration of the TeV scale, notably in detailed studies of the
Higgs boson, assuming it is light enough, and of any accessible new
physics appearing at the electroweak scale, such as
supersymmetry~\cite{LCphys}. A LC could also provide important and
distinctive indirect tests of unification ideas~\cite{BPZ}, and also
explore physics in extra dimensions, if they open up at a low enough
energy scale.

These are strong arguments, which need to be developed at several
different levels. Beyond the community assembled here, are they strong
enough to convince the world-wide particle-physics community that it
should unite around a consensus that the next major global project after
the LHC should be a LC? And are they strong enough to convince circles in
the the world outside particle physics, notably other physicists, other
scientists, funding agencies and politicians?

They may be easier to persuade once (if) the Higgs boson is discovered.
The Higgs mass probability distribution obtained by combining direct and
indirect information suggests that the Higgs boson may lie quite close to
the present experimental lower limit of 114.4 GeV, with a 95$\%$
confidence-level upper limit in the Standard Model of about 260
GeV~\cite{LEPEWWG}. Unfortunately, LEP was not able to discover the Higgs
boson, and the hint found at the end of 2000 has finally diminished to
below two standard deviations. CDF and D0 may be able to find some
evidence before the start-up of the LHC, but the LHC should be able to
make a 5-$\sigma$ discovery with 10~fb$^{-1}$ of data, which should be
obtainable in 2008~\cite{LHCH}.

\begin{figure}[htb]
\includegraphics[width=.48\textwidth]{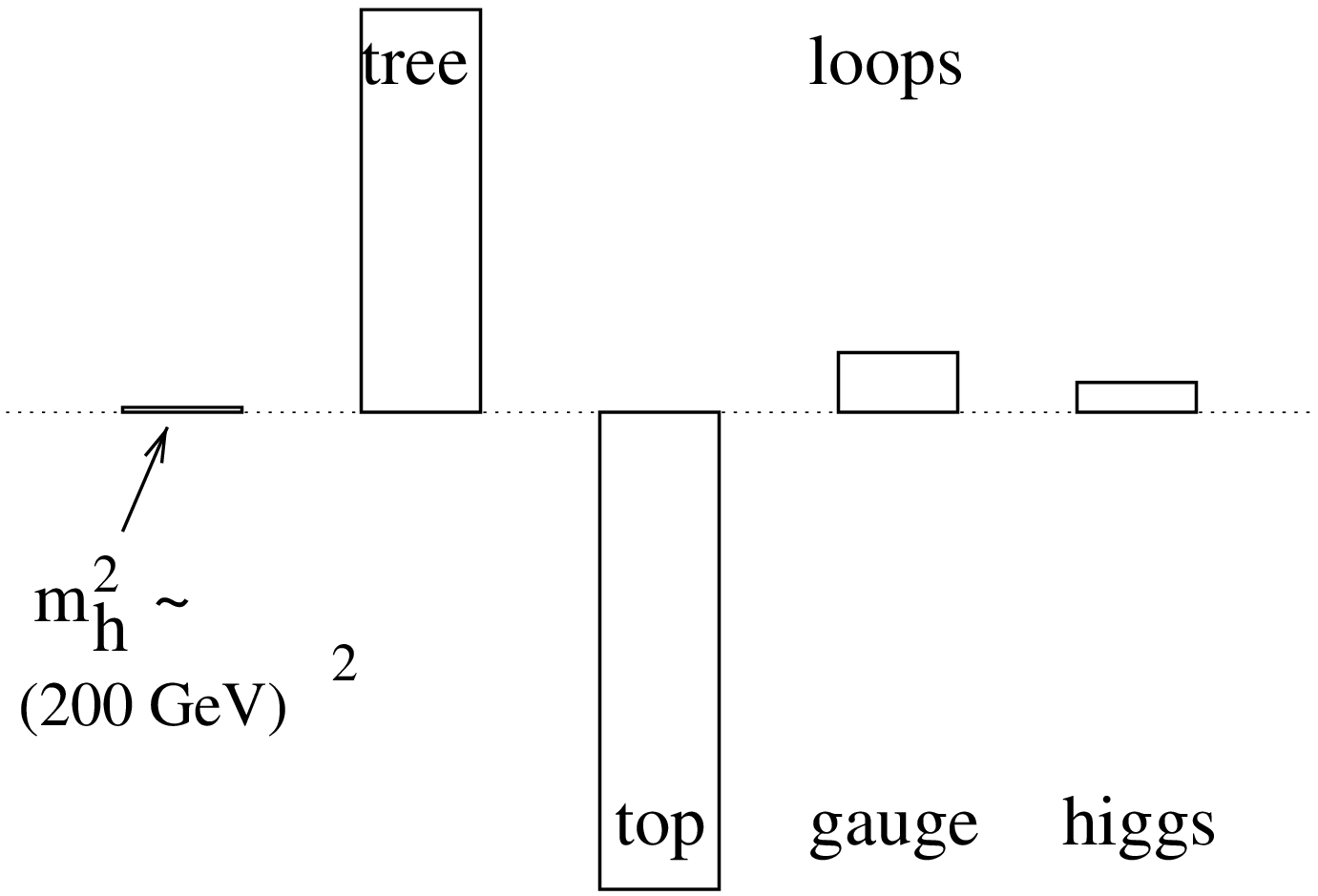}
\includegraphics[width=.48\textwidth]{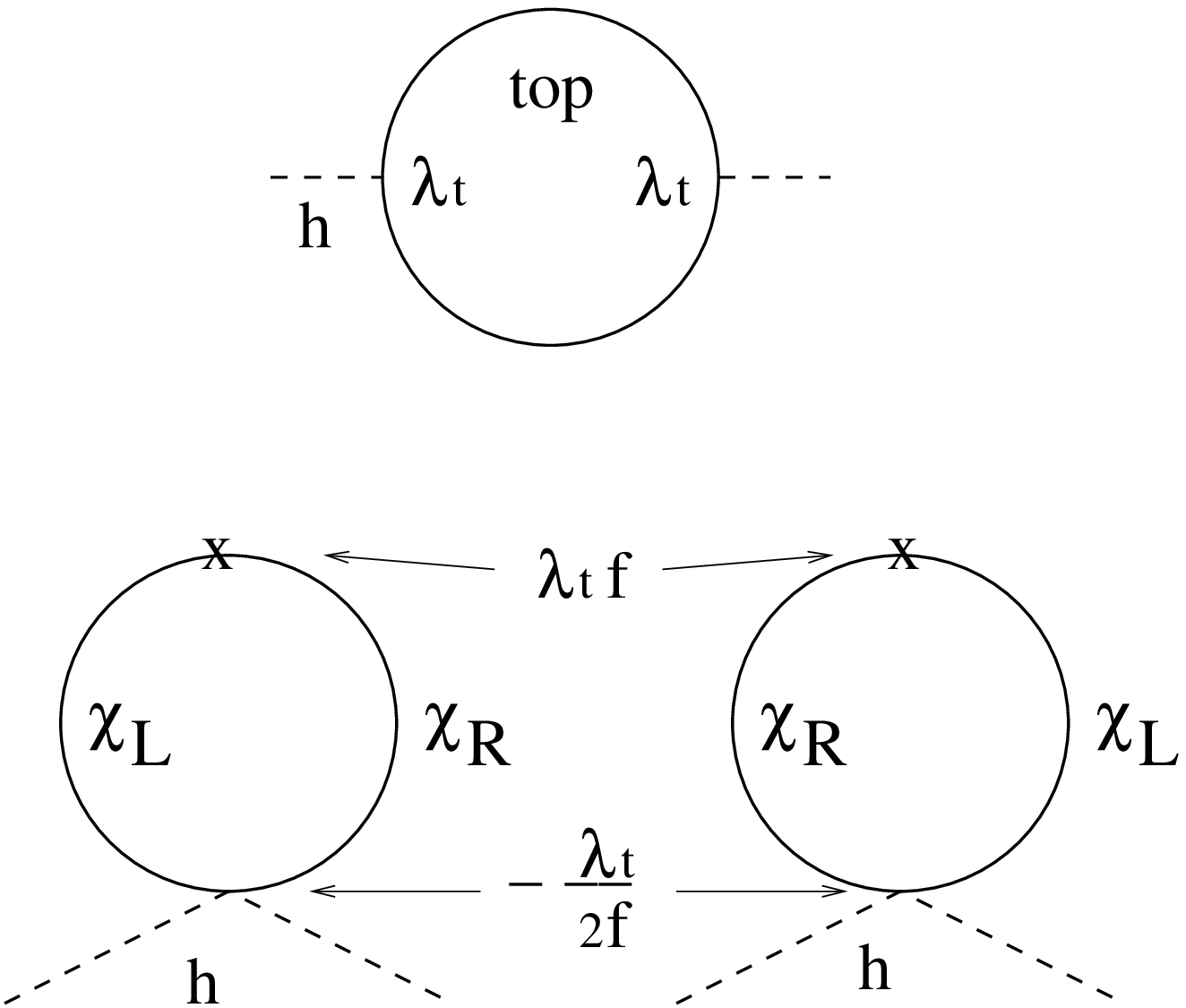}
\caption{(a) If the Standard Model Higgs boson weighs around 200~GeV, the 
top-quark loop contribution to its physical mass (calculated here with a loop 
momentum cutoff of 10~TeV) must cancel delicately against 
the tree-level contribution. (b) In `Little 
Higgs' models, the top-quark loop is cancelled by 
loops containing a heavier charge-2/3 quark~\protect\cite{Schmaltz}.} 
\label{fig:cancel}
\end{figure}

\section{Theorists are Getting Cold Feet}

With make-or-break time for discovery of the Higgs boson at the LHC
looming closer on the horizon - and supersymmetry, if it exists at the
electroweak scale - at least some theorists are getting cold feet,
exploring various avenues for avoiding a light Higgs boson and/or
supersymmetry.

Some used to question whether {\it the Higgs boson might be 
composite}~\cite{HS},
but this possibility seems to be inconsistent with precision electroweak
data~\cite{technicolour}, and is not discussed further here. However,
questions have been raised about the interpretation of the electroweak
data: {\it are the different measurements consistent, or should some be
discarded?} If so~\cite{Chan}, what would happen to the upper limit on
$m_h$ and would there be other signatures of new physics? Even if one
accepts the electroweak data at face value, is the renormalizable
Lagrangian of the Standard Model adequate for describing them, or {\it
should one supplement the Lagrangian with higher-dimensional operators?}
If one includes such higher-dimensional operators, are credible corridors
to higher Higgs masses opened up~\cite{Barb}? Even if this is not the
case, and the Higgs is relatively light, {\it is supersymmetry the only
mechanism for avoiding the fine-tuning problem?} New alternatives are
provided by `Little Higgs' models, in which there are one-loop
cancellations with an extra top-like quark, gauge bosons and additional
Higgs fields~\cite{Schmaltz}. Finally, {\it is it really established that
a Higgs boson must exist?} This question has been asked again within a new
wave of Higgsless models, which must deal with strong $WW$ scattering and
the ensuing implications for precision electroweak
observables~\cite{Higgsless}.

We now discuss some more aspects of these scenarios for avoiding the
conventional light Higgs/supersymmetry scenario.

\subsection{Heretical Interpretations of the Electroweak Data}

It is notorious that the two most precise measurements at the $Z$ peak,
namely the asymmetries measured with leptons and hadrons favour different
values of $m_h$, around 50 and 400 GeV, respectively~\cite{LEPEWWG}.
Perhaps this discrepancy is just a statistical fluctuation, or perhaps we
do not understand hadronic systematics as well as we think~\cite{Chan}?
Another anomaly is exhibited by the NuTeV data on deep-inelastic $\nu - N$
scattering~\cite{NuTeV}, which may be easier to explain away as due to our
lack of understanding of hadronic effects. On the other hand, if either
the lepton/hadron discrepancy or the $\nu - N$ anomaly is genuine, there
may be new physics at the electroweak scale. In this case there would be
no firm basis for the prediction of a light Higgs boson, which is based on
a Standard Model fit~\cite{Chan}. Unfortunately, it is unclear how the $Z$
peak discrepancy could be cleared up soon, whereas NOMAD may soon cast
some light on the NuTeV anomaly.

\subsection{Higher-Order Operators}

If one regards the Standard Model simply as an effective low-energy theory,
one should expect the renormalizable
dimension-four interactions it contains to be supplemented by
higher-dimensional operators of the general form~\cite{Barb}:
$$
{\cal L}_{eff} = {\cal L}_{SM} + \sum_i \; \frac{c_i}{\Lambda_i^p} \; 
{\cal O}_i^{4+p}.
$$
A global fit to the precision electroweak data
suggests that, if the Higgs is indeed light,
the coefficients of these additional interactions are small: $\Lambda_i 
\sim 10$~TeV for $c_i = \pm 1$. The `little hierarchy' problem is to 
understand why this should be the case~\cite{littlehier}.

\begin{figure}[htb]
\begin{center}
\includegraphics[width=12cm]{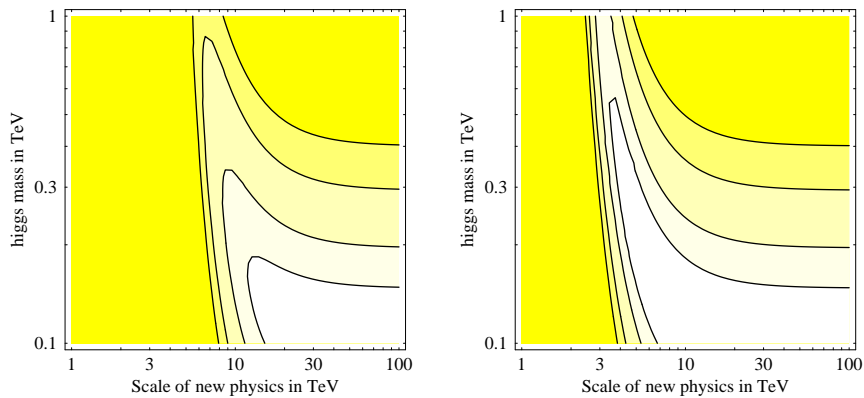}
\caption{If suitable non-renormalizable operators $O_{WB}$ or $O_H$ are 
included in the 
global electroweak fit with $c_i = -1$, corridors of parameter space 
leading to a large 
mass for the Higgs boson may be opened up~\protect\cite{Barb}.}
\label{fig:corridor}
\end{center}
\end{figure}

However, conspiracies are in principle possible, enabling $m_H$ to be
large, even if one takes the precision electroweak data at face value.
Examples are shown in Fig.~\ref{fig:corridor}, where one sees corridors of
the allowed parameter space extending up to a heavy Higgs
mass~\cite{Barb}. Any theory beyond the Standard Model must link the value
of $m_H$ and the coefficients of these higher-dimensional effective
operators in some way. A theory that predicts a heavy Higgs boson but
remains consistent with the precision electroweak data should predict a
correlation of the type seen in Fig.~\ref{fig:corridor}. At the moment,
this may seem unnatural to us, but Nature may know better.

For the moment, we should not discard the possibility of a heavy Higgs
boson. However, Fig.~\ref{fig:corridor} suggests that, if this is the
case, there would be observable effects due to higher-dimensional
effective operators that could be measured at a LC~\cite{LCphys}.

\subsection{Little Higgs Models}

These models invoke a different mechanism for enforcing
the loop cancellations needed to keep a
light Higgs boson light. The strategy is to embed the Standard Model in a
larger gauge group which is then broken
spontaneously down to the Standard Model~\cite{Schmaltz}. The Higgs boson 
appears as a
pseudo-Goldstone boson, which guarantees
that it is light before the loop effects kick in. Generally, the top-quark 
loop
contribution to the Higgs mass-squared has
the general form
$$
\delta m^2_{H, top}(SM) \sim (115 \; \rm{GeV} )^2 \;
\left( \frac{\Lambda}{400 \rm{GeV}} \right )^2
$$
As illustrated in Fig.~\ref{fig:cancel}, in Little Higgs models this is 
cancelled by the
loop contribution due to a new heavy
top-like quark $T$, leaving
$$
\delta m^2_{H, top}(LH) \sim
\frac{6G_F m^2_t}{\sqrt{2 \pi^2}} \, m^2_T \, \log \frac{\Lambda}{m_T}.
$$
Analogously, to cancel the loop divergences associated with the gauge
bosons and the Higgs boson of the Standard
Model, Little Higgs models contain new gauge bosons and Higgs bosons.

\begin{figure}[htb]
\begin{center}
\includegraphics[width=6cm,angle=-90]{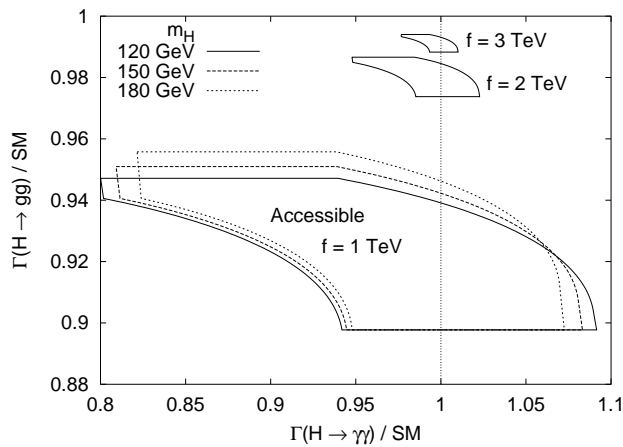}
\caption{The rates for Higgs decays into $\gamma 
\gamma$ and gluon pairs are sensitive to the scale $f$ in little Higgs 
models, enabling precision measurements at a LC to distinguish them from 
a Standard Model Higgs boson at the same mass~\protect\cite{LHCLC}.} 
\label{fig:littlehiggs}
\end{center}
\end{figure}

The net result is a spectrum containing a relatively light Higgs boson and
other new particles that may be somewhat heavier:
\begin{equation}
M_T < 2  \rm {TeV} (\frac{m_h}{200~\rm{GeV}})^2, \;
M_{W'} < 6  \rm {TeV} (\frac{m_h}{200~\rm{GeV}})^2, \;
M_{H^{++}} < 10  \rm {TeV} .
\end{equation}
In addition, there should be more
physics at some energy scale above 10 TeV, for the ultra-violet completion
of the theory. Some of these new particles should be accessible to the 
LHC~\cite{LHLHC} and, if the 
new particles predicted in such models are within the reach for
direct production at a LC, it will be able to explore them in detail. Even
if not, a LC can probe such a model via careful studies of its light Higgs
boson, e.g., by measuring accurately its decays into $\gamma \gamma$ and
gluon pairs~\cite{LHCLC}, as seen in Fig.~\ref{fig:littlehiggs}.

\subsection{Higgsless Models}

The most radical alternative to the Higgs sector in the Standard Model is
offered by Higgsless models, which were originally formulated in the
conventional four dimensions~\cite{Higgsless}. Inverting the usual
argument that $WW$ scattering would violate unitarity if there were no
Higgs boson, one would expect such Higgsless models to exhibit strong $WW$
scattering at the TeV scale. This is likely, {\it a priori}, to be
incompatible with the precision data. The second wave of Higgsless models
addressed this problem by adding an extra dimension, and postulating
boundary conditions that break the electroweak symmetry~\cite{Higgsless5}.
These extra-dimensional variants are able to delay the onset of strong
$WW$ scattering to about 10~TeV, and the simplest variants exhibit a
forest of Kaluza-Klein excitations with masses starting above 300~GeV.
However, compatibility with the precision electroweak data is still an
issue for such models, motivating epicyclic variants with a warped extra
dimension and special brane kinetic terms~\cite{Rizzo}.

Clearly, if the lightest Kaluza-Klein modes do have masses around 300 GeV,
they would provide directly a cornucopia for a LC. Additionally, the
sort of massive resonance predicted in models with strong $WW$ scattering
might be detectable indirectly at a LC.

\section{Measuring the Properties of a Light Higgs Boson}

The capabilities of a sub-TeV LC for precision measurements of the
branching ratios for a light Higgs boson into modes such as ${\bar b} b,
\tau \tau, gg, {\bar c}c, WW$ and $\gamma \gamma $ are well
documented~\cite{LCphys}, and some new studies have recently become
available. For example, the capabilities of the LHC and a LC for measuring
the top-Higgs coupling have recently been evaluated in the context of the
joint LHC/LC study~\cite{LHCLC}. 

\begin{figure}[htb]
\begin{center}
\includegraphics[width=.48\textwidth]{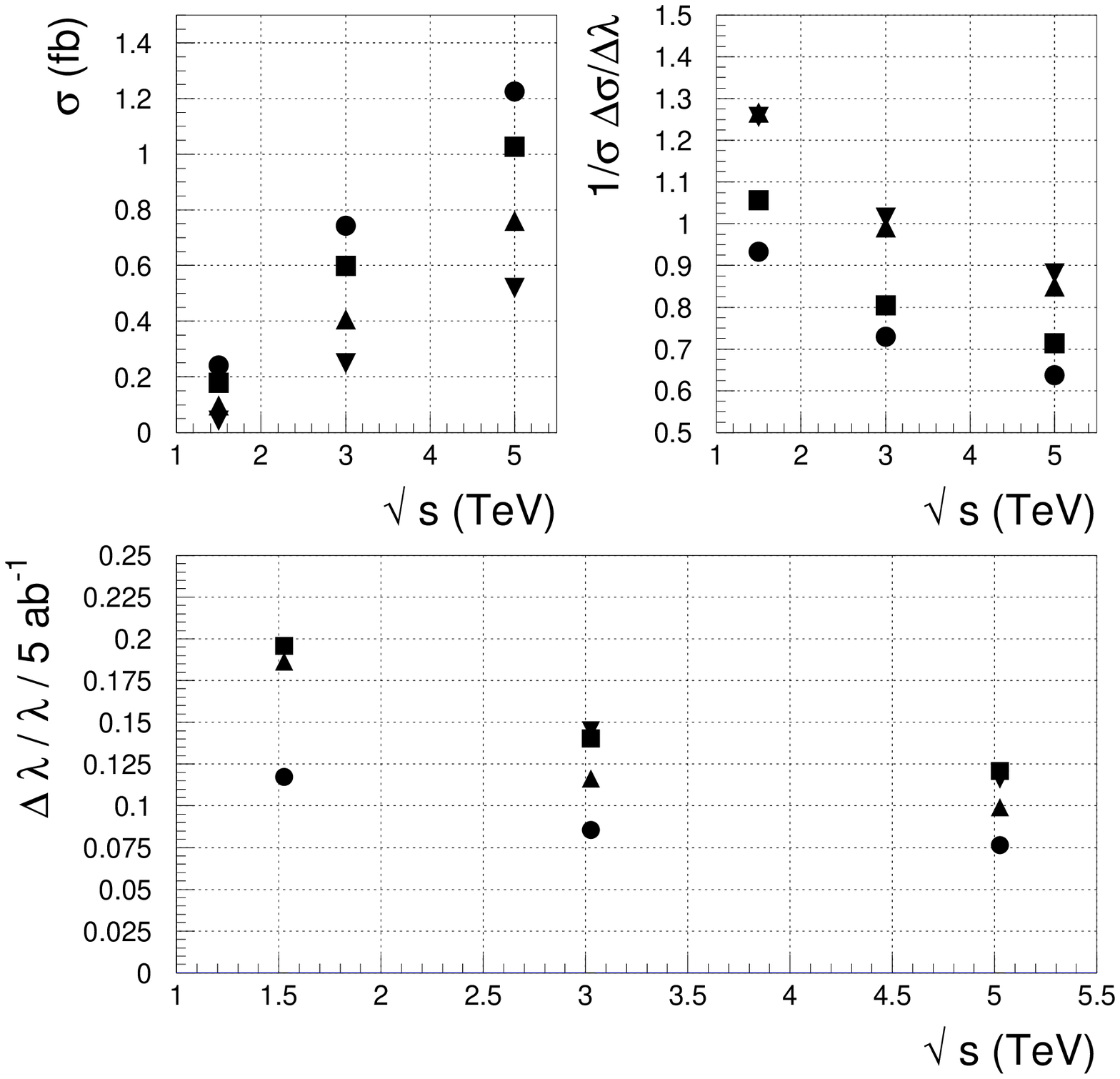}
\includegraphics[width=.48\textwidth]{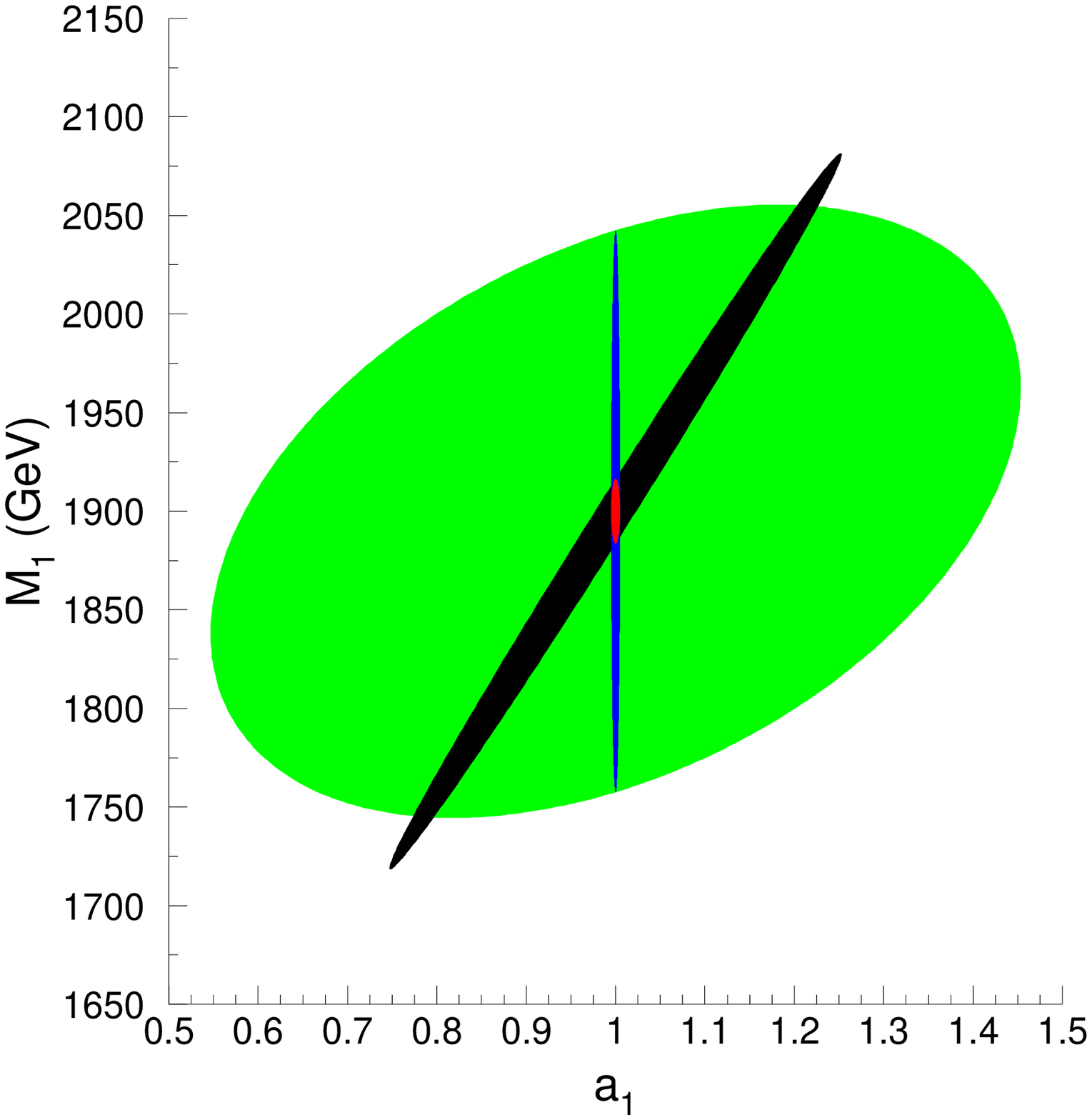}
\caption{(a) Comparison of the accuracies with which the triple-Higgs 
coupling could be measured at LCs with different centre-of-mass 
energies~\protect\cite{CLICphys}, and (b) comparison of possible 
measurements of a massive $WW$ resonance 
at the LHC (broad ellipse) and a 500-GeV LC (narrow ellipse), showing also 
the potential impact of knowledge of the coupling parameter 
$a_1$~\protect\cite{LHCLC}.}
\label{fig:HHH}
\end{center}
\end{figure}

It has also been realized recently that a
higher-energy LC has certain advantages for precision measurements, even
of a light Higgs boson, due mainly to the much larger cross sections for
Higgs production at multi-TeV energies. For example, one can measure
accurately rare decay modes, such as $H \to \mu \mu$ for $m_H = 120$ GeV
and $H \to {\bar b} b$ for $m_H = 180$ GeV~\cite{CLICphys}.
Another topic where a higher-energy LC has an advantage is in measuring
the Higgs self-couplings. It is well known that the trilinear Higgs
coupling of a light Higgs boson can be measured at a low-energy
LC~\cite{LCphys}, and it has recently been shown that this might be
possible for a heavier Higgs boson at the luminosity upgrade of the LHC,
the SLHC~\cite{LHCLC}. A study has also been made of the measurement of
the effective Higgs potential using a 3-TeV LC. This would have a much
larger cross section for $HH$ pair production than a sub-TeV LC, enabling
the accuracy in the measurement of the HHH coupling to be improved for all
masses between 120 and 240 GeV, as seen in
Fig.~\ref{fig:HHH}(a)~\cite{CLICphys}.

Strong $WW$ scattering may be parameterized by effective higher-order
gauge-boson couplings that appear at the quartic level. These can be
measured in $WW, WZ$ and $ZZ$ final states at the LHC, and (more
accurately) in $WW$ and $ZZ$ final states at a LC. Going beyond the
effective `low-energy' interactions, one or more $WW$ resonances may
appear. The first hint of a $WW$ resonance may be given by form-factor
measurements, and a 500-GeV LC would be able to probe the existence of a
$\rho$-like resonance far beyond its direct energy reach. The parameters
of such a resonance might be measured first at the LHC, but they could be
measured more precisely at a 500-GeV LC, as seen in 
Fig.~\ref{fig:HHH}(b)~\cite{LHCLC}.

\begin{figure}[htb]
\includegraphics[width=.48\textwidth]{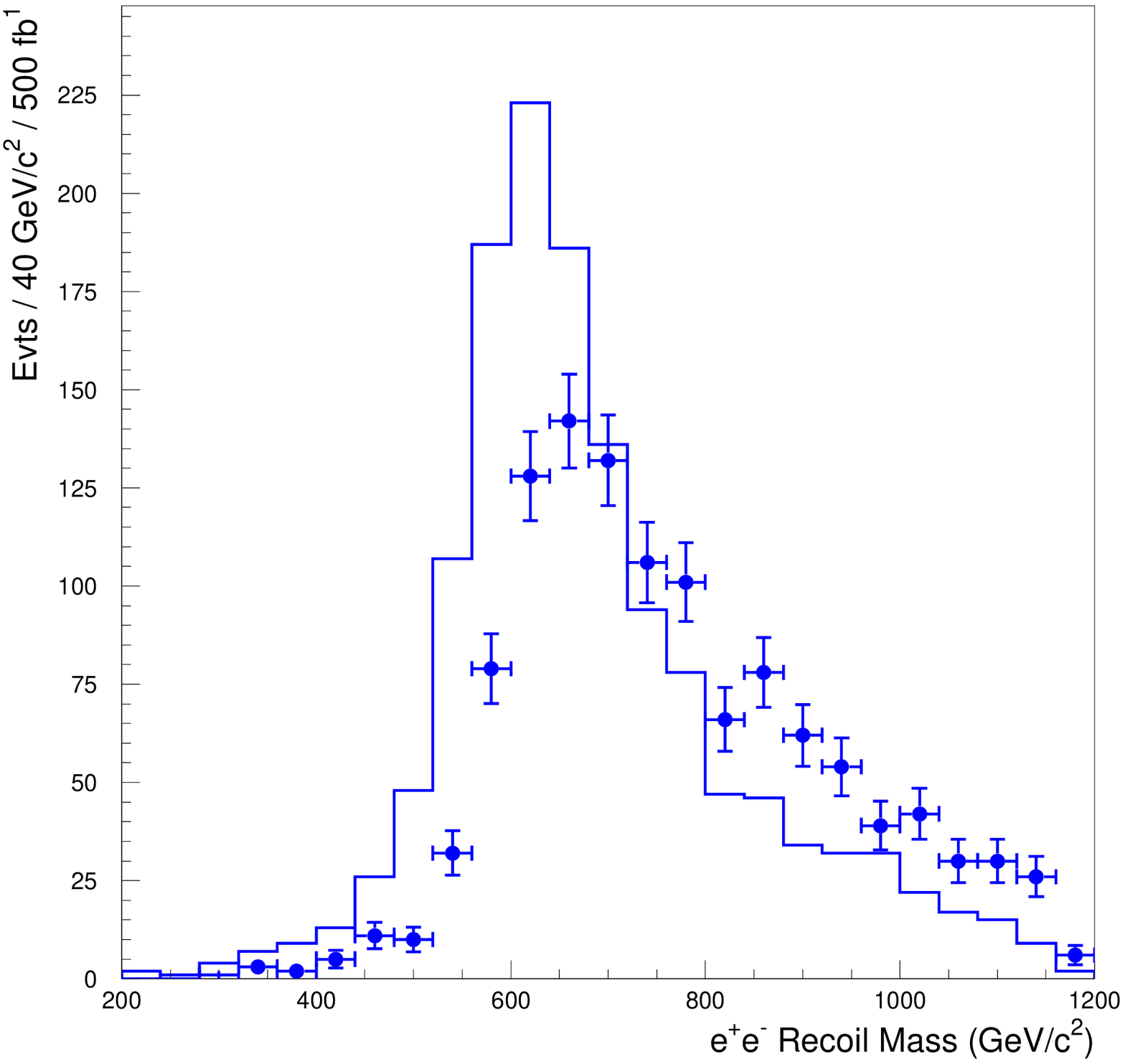}
\includegraphics[width=.48\textwidth]{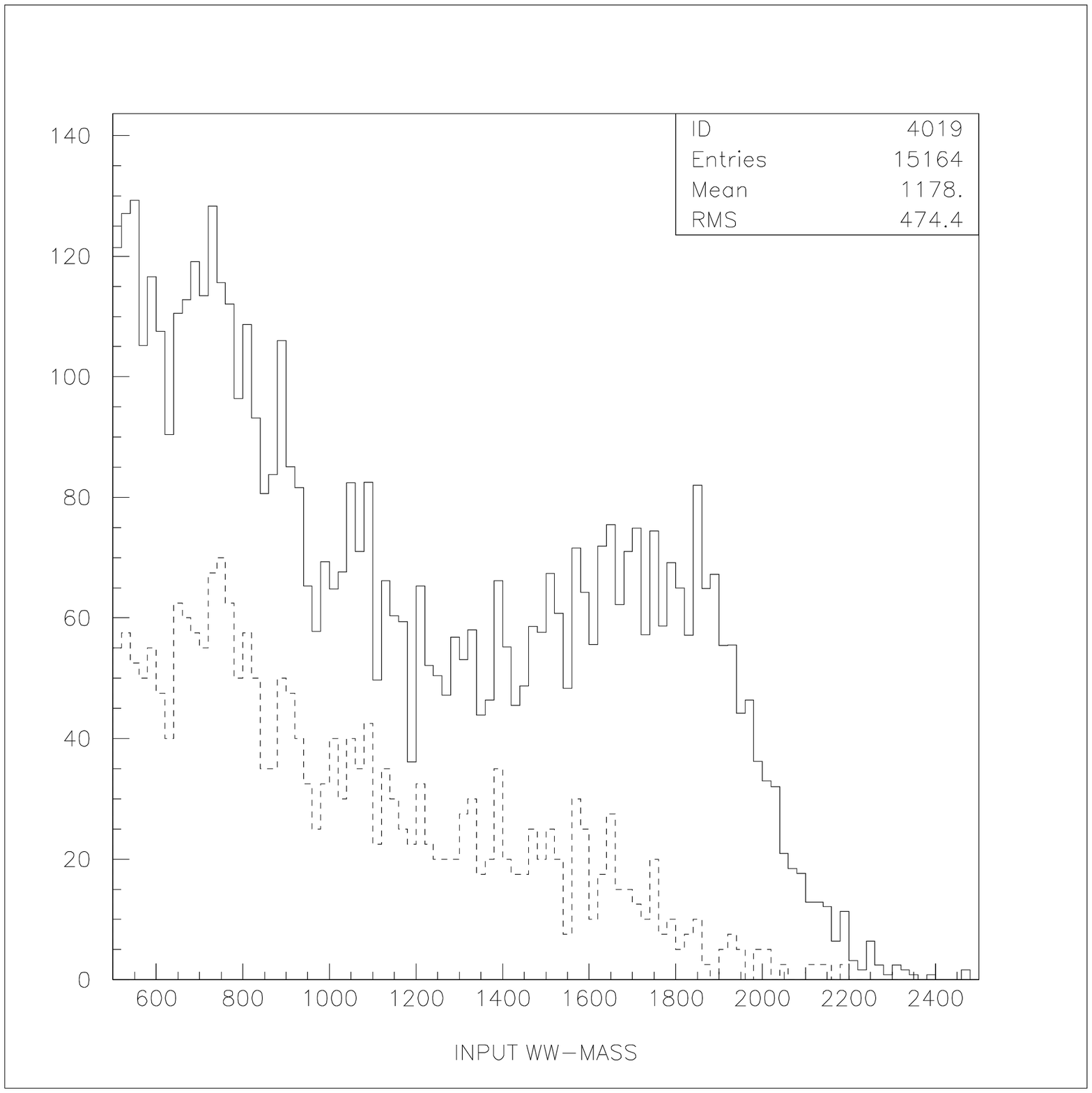}
\caption{Examples of possible signals for the direct production of 
massive Higgs resonances in (a) $e^+ e^- \to H e^+ e^-$ and (b) $e^+ e^- 
\to H \nu {\bar \nu}$ at a 3-TeV LC~\protect\cite{CLICphys}.} 
\label{fig:heavyH}
\end{figure}

Such a $WW$ resonance might be observable directly for the first time at a
multi-TeV LC. The channel $ee \to H ee$ could be used to establish its
existence beyond any doubt if it weighs $<$ 1 TeV, , as seen in
Fig.~\ref{fig:heavyH}(a), and one could find a resonance in strong WW
scattering via the $e^+ e^- \to H \nu {\bar \nu}$ channel even if it
weighs $>$ 1 TeV, as seen in Fig.~\ref{fig:heavyH}(b)~\cite{CLICphys}.

\section{Other New Physics at the Electroweak Scale}

If the Higgs boson is light, and in particular if it is very close to the
direct search limit, the effective Higgs potential is in danger of being
destabilized by the loop corrections due to the top quark. These tend to
turn the effective Higgs potential negative far below the Planck scale,
and the loop corrections due to a light Higgs boson would be insufficient
to counteract them. These should be supplemented by new physics appearing
at a relatively low energy scale, for which the best candidate may be
supersymmetry~\cite{ER}. This argument is independent of the primary
motivation for expecting supersymmetry at accessible energies below about
a TeV, namely the hierarchy problem. Other reasons for
liking supersymmetry include the fact that it enables the gauge couplings
to unify (though there are other {\it ad hoc} fixes for
this), the prediction of a low Higgs mass~\cite{susyH}, and a natural
candidate for the cold dark matter advocated by astrophysicists and
cosmologists~\cite{EHNOS}.

To my mind, none of the alternatives currently available on the market --
extra dimensions, little Higgs models, Higgsless models, etc.-- are as
satisfactory as supersymmetry. This is not to say that supersymmetry is
completely satisfactory itself - for example, the mechanism and scale of
supersymmetry breaking are obscure and supersymmetry does not, by itself,
explain the magnitude of the electroweak scale, it merely stabilizes it.
However, supersymmetry often appears as a component in these alternative
scenarios, for example to stabilize the scales of extra dimensions.

\section{Studies of Supersymmetry}

The minimal supersymmetric extension of the Standard Model (MSSM) contains
over 100 parameters, mostly in the parameters that break supersymmetry via
masses $m_0$ for the spin-0 supersymmetric partners of the Standard Model
fermions, masses $m_{1/2}$ for the spin-1/2 supersymmetric partners of the
Standard Model bosons and trilinear soft supersymmetry-breaking parameters
$A$. In order to visualize the parameter space, one often makes
simplifying assumptions about these parameters, and it is popular to
assume that the parameters $m_0, m_{1/2}$ and $A$ are universal for the
different sparticle types, in the so-called constrained MSSM (CMSSM).

\begin{figure}[htb]
\begin{center}
\includegraphics[width=.48\textwidth]{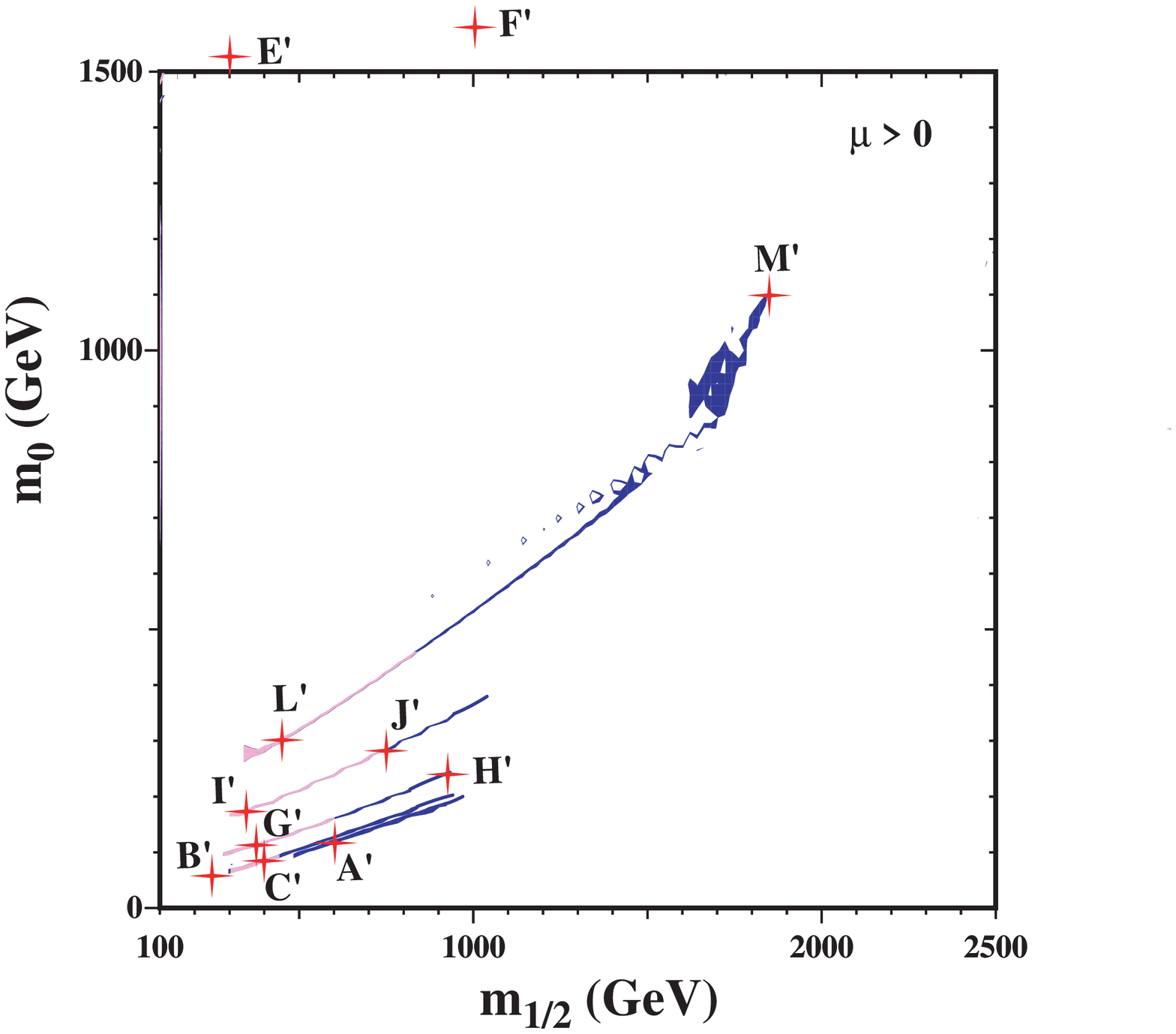}
\includegraphics[width=.48\textwidth]{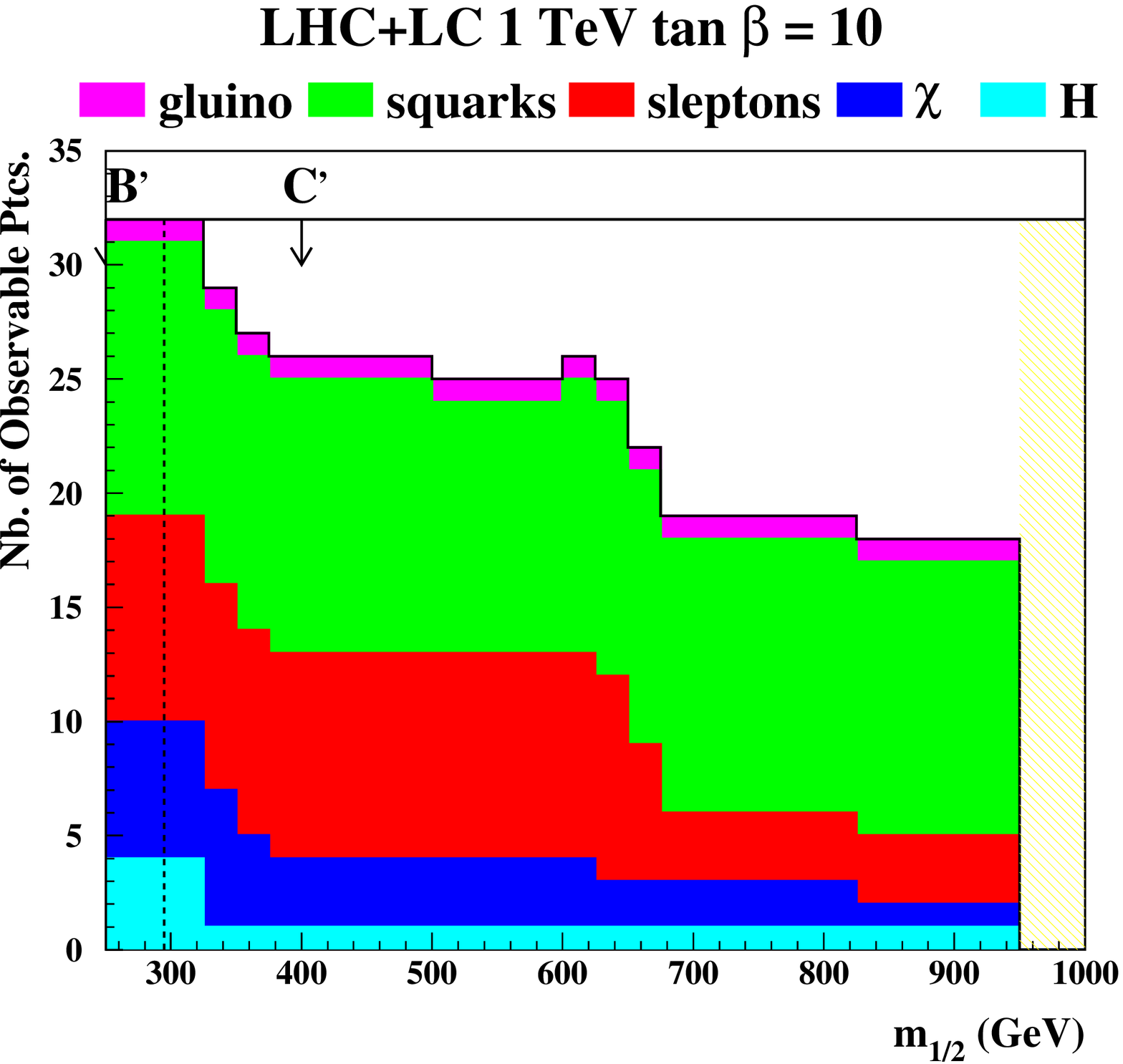}
\end{center}
\caption{(a) The strips of CMSSM parameter space allowed by WMAP and 
other constraints for $\mu > 0$ and different choices of $\tan \beta$, 
with candidate benchmark points indicated, and (b) the numbers of types 
of supersymmetric particles detectable along the WMAP strip for $\tan 
\beta = 10$, combining the LHC and a 1-TeV LC~\protect\cite{Bench2}.} 
\label{fig:WMAPline} 
\end{figure}

\begin{figure}[htb]
\includegraphics[width=.48\textwidth]{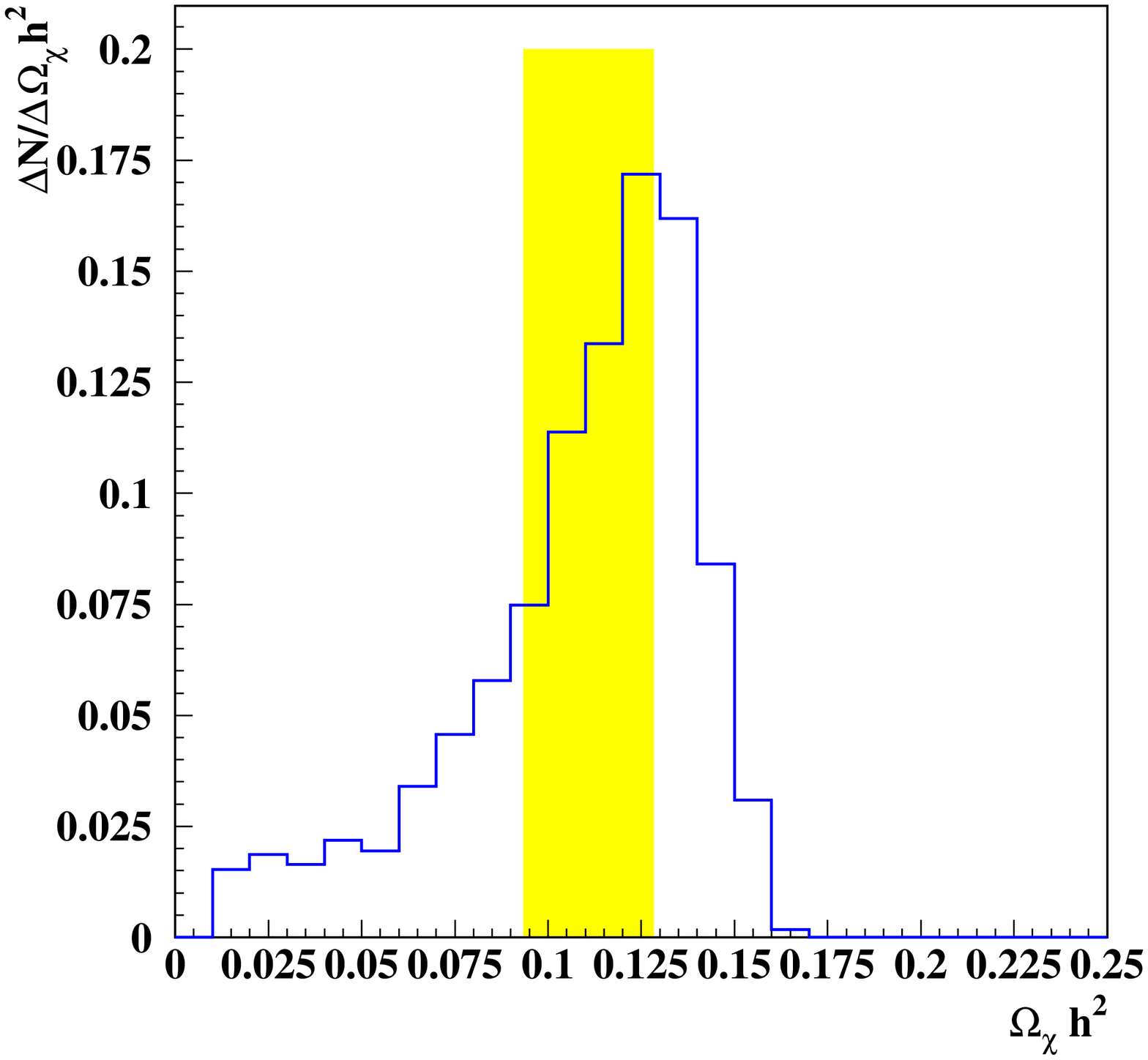}
\includegraphics[width=.48\textwidth]{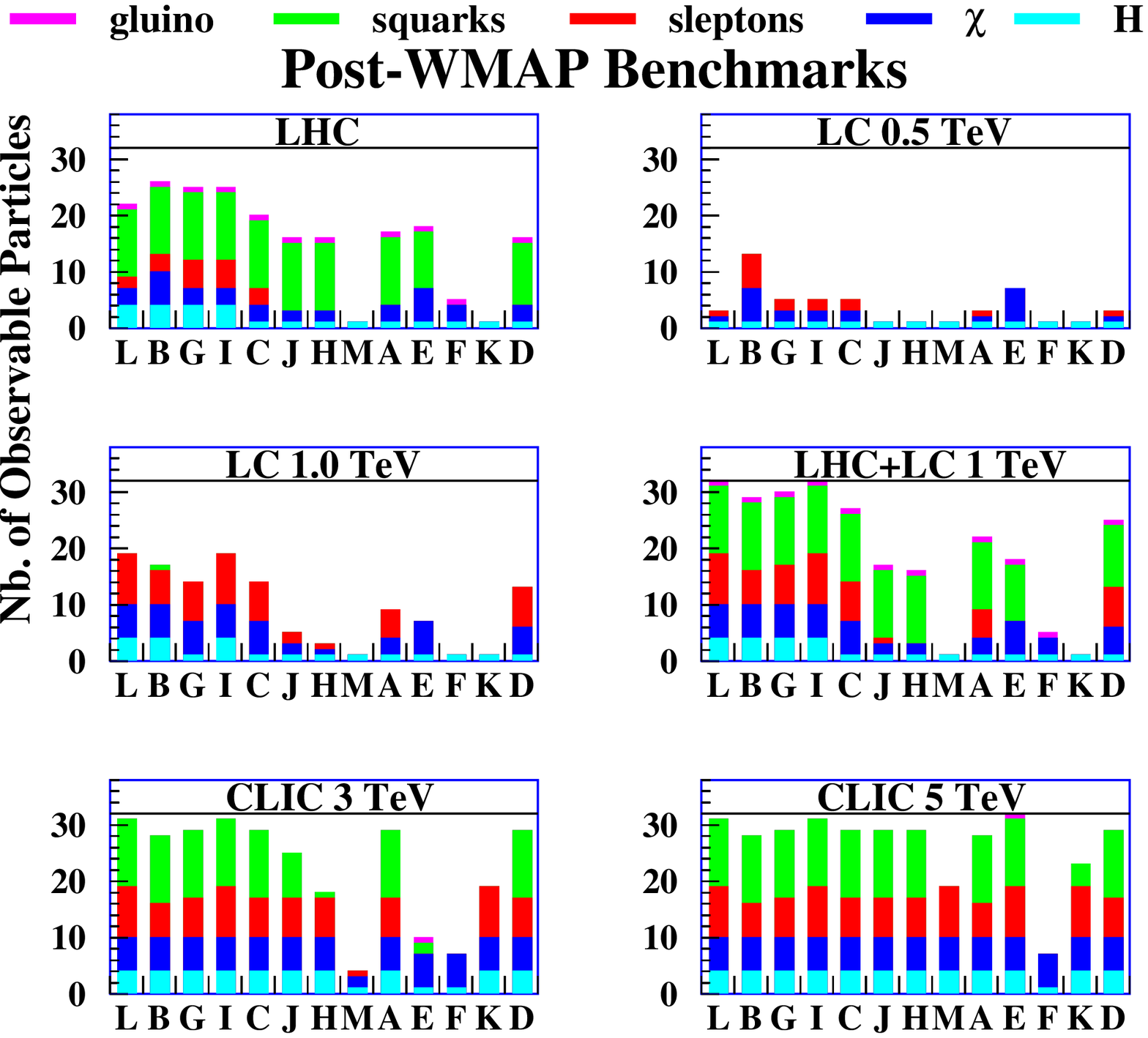}
\caption{(a) A comparison of the accuracy with which the relic neutralino 
density could be estimated using LHC measurements with the range allowed 
by WMAP, assuming benchmark scenario B of Fig.~\protect\ref{fig:WMAPline}, 
and (b) the numbers of types of supersymmetric particles detectable 
at various colliders in the different benchmark scenarios indicated in 
Fig.~\protect\ref{fig:WMAPline}(a)~\protect\cite{Bench2}.} 
\label{fig:Omega}
\end{figure}

The regions of CMSSM parameter space allowed by the accelerator and dark
matter constraints, particularly in the latest version after the WMAP
data, are typically narrow lines, as seen in
Fig.~\ref{fig:WMAPline}(a)~\cite{Bench2,earlierbench}. One may then study
the capabilities of different accelerators to make measurements as one
varies the CMSSM parameters along one of these WMAP lines, as exemplified
in Fig.~\ref{fig:WMAPline}(b), or one may choose to study in more detail
benchmark points located at specific places along these lines, as
indicated in Fig.~\ref{fig:WMAPline}(a). Fig.~\ref{fig:WMAPline}(b)
displays the numbers of different sparticle species that would be
detectable at the LHC and/or a 1-TeV LC as one varies parameters along the
WMAP line for $\tan \beta = 10$ and $\mu > 0$~\cite{Bench2}. The LHC
measurements would enable one to calculate the relic LSP density and check
whether it falls within the WMAP range. Fig.~\ref{fig:Omega}(a) shows the
result of one such calculation based on realistic errors in such LHC
measurements, assuming the parameters of one specific benchmark
point~\cite{Bench2}. The error is already comparable with the WMAP
uncertainty, and could be refined significantly with the aid of LC
measurements~\cite{Richard}.

Fig.~\ref{fig:Omega}(b) compares the numbers of different sparticle
species that could be detected at the LHC and linear colliders of
different energies~\cite{earlierBench,Bench2}. We see that the LHC and LC
have complementary scapabilities, in that the LC can observe many types of
weakly-interacting sparticle that would be invisible at the LHC - as long
as the LC centre-of-mass energy is above its production threshold. In many
cases, several sparticles would be seen at a 500-GeV LC, more would be
seen at a 1000-GeV LC, and others would require an even higher
centre-of-mass energy. High-precision sparticle measurements could be made
via the positions of edges in dilepton spectra at the LHC, followed by
threshold measurements and final-state lepton spectra at a LC. In this and
other examples, the LC measurements would provide considerable added value
for the determination of sparticle masses and the underlying CMSSM
parameters~\cite{LHCLC}, making possible crucial tests of our ideas about
grand unification of sparticle masses as well as gauge
couplings~\cite{BPZ}.

\begin{figure}[htb]
\begin{center}
\includegraphics[width=.48\textwidth]{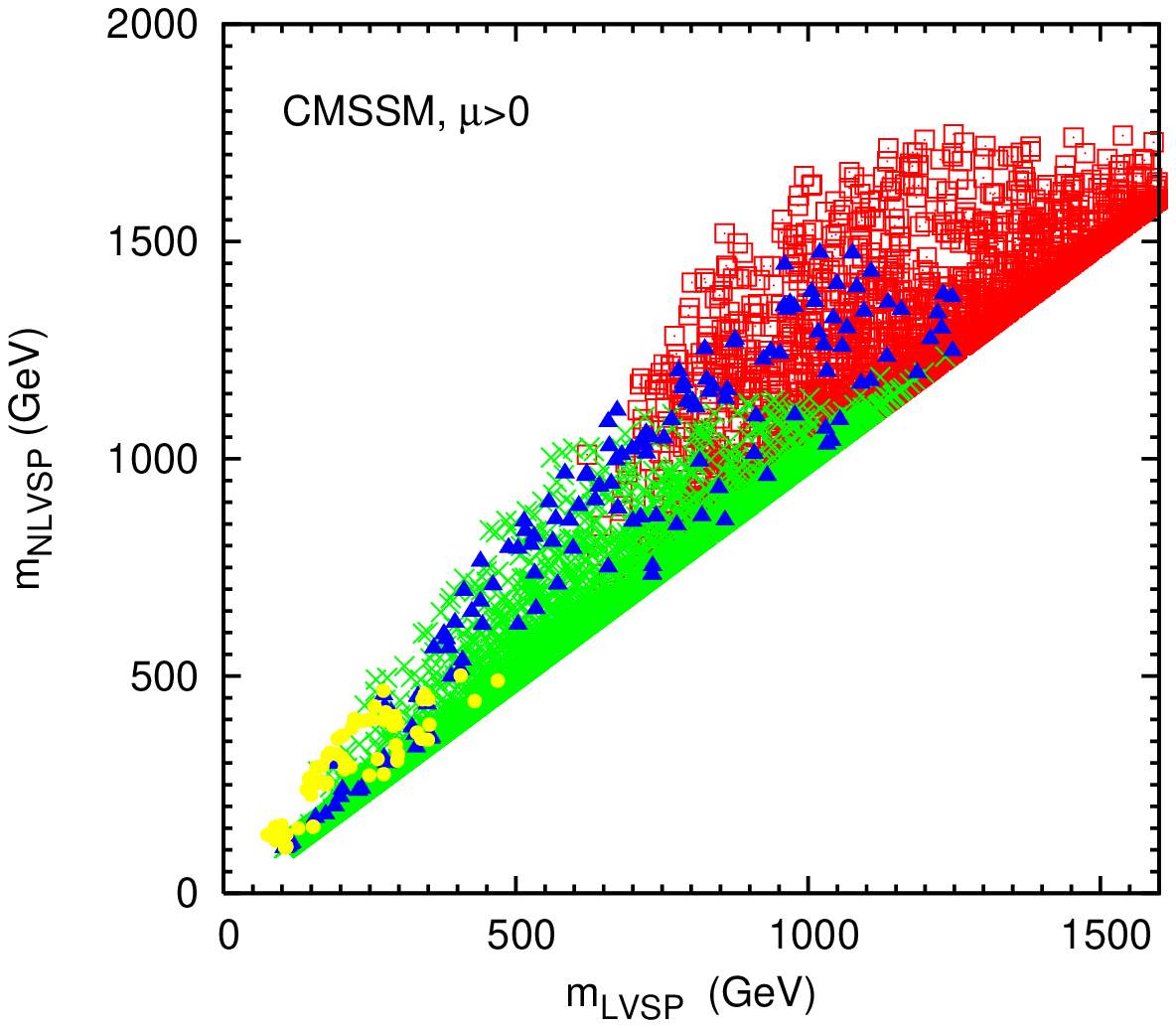}
\includegraphics[width=.48\textwidth]{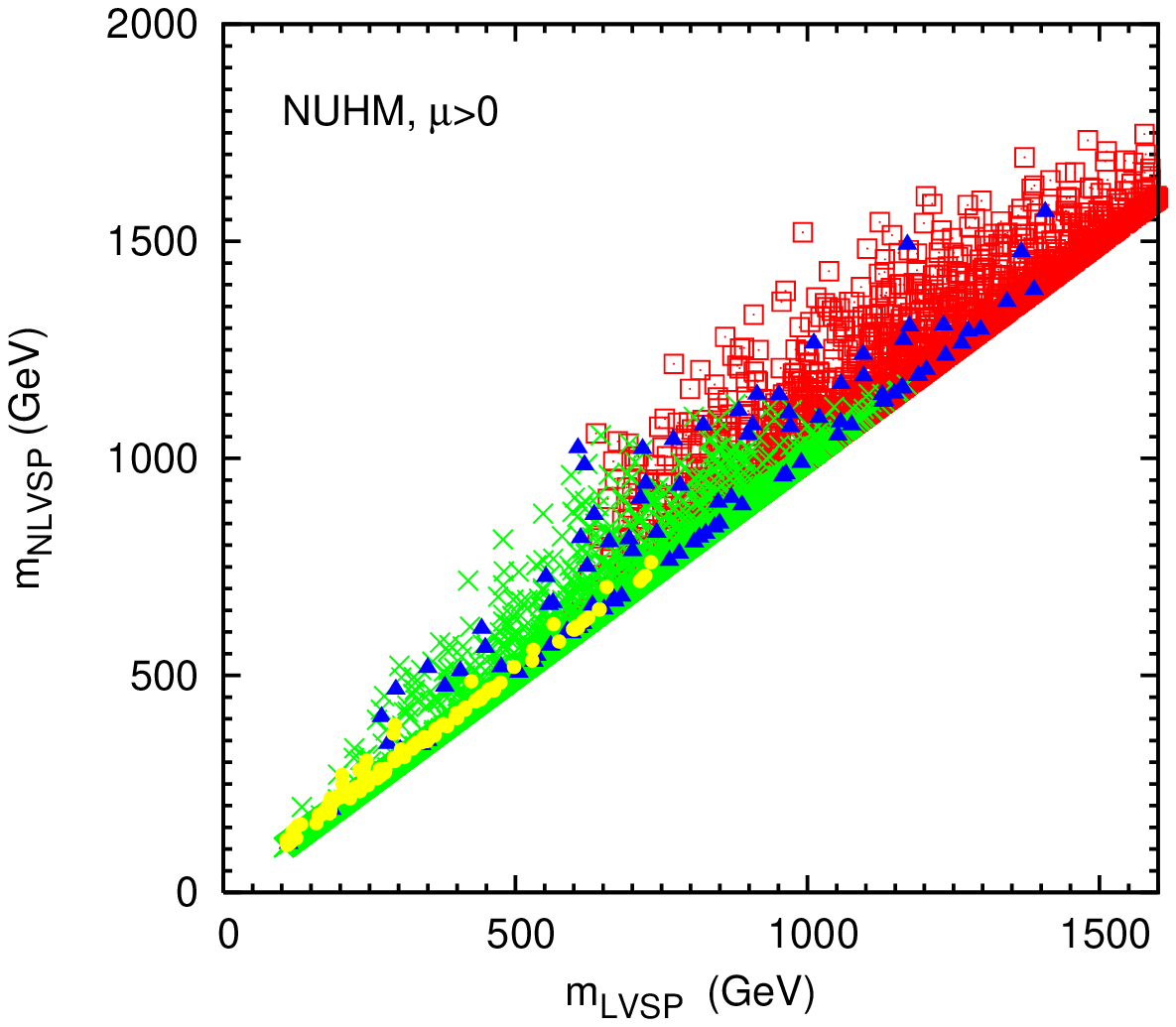}
\includegraphics[width=.48\textwidth]{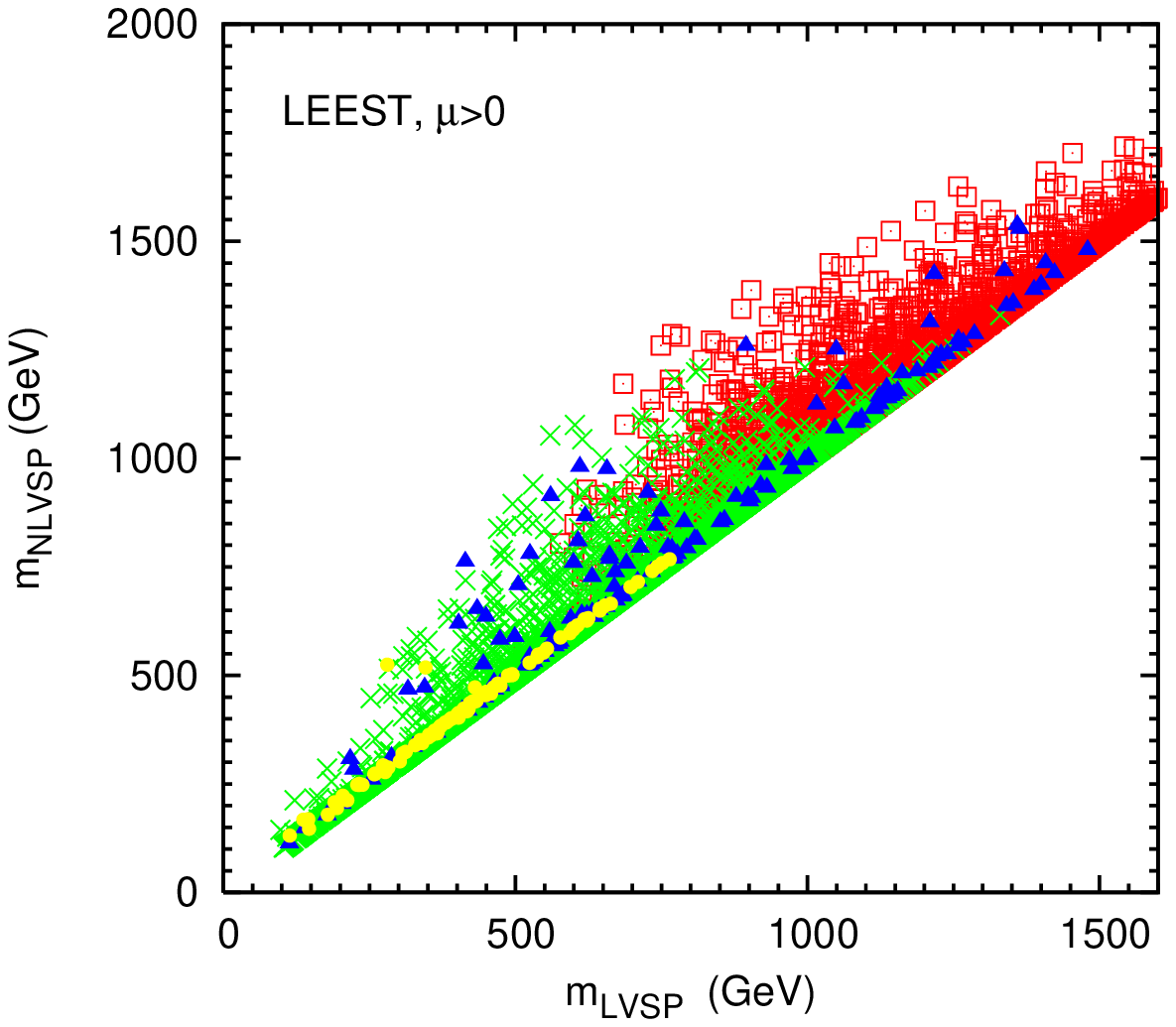}
\includegraphics[width=.48\textwidth]{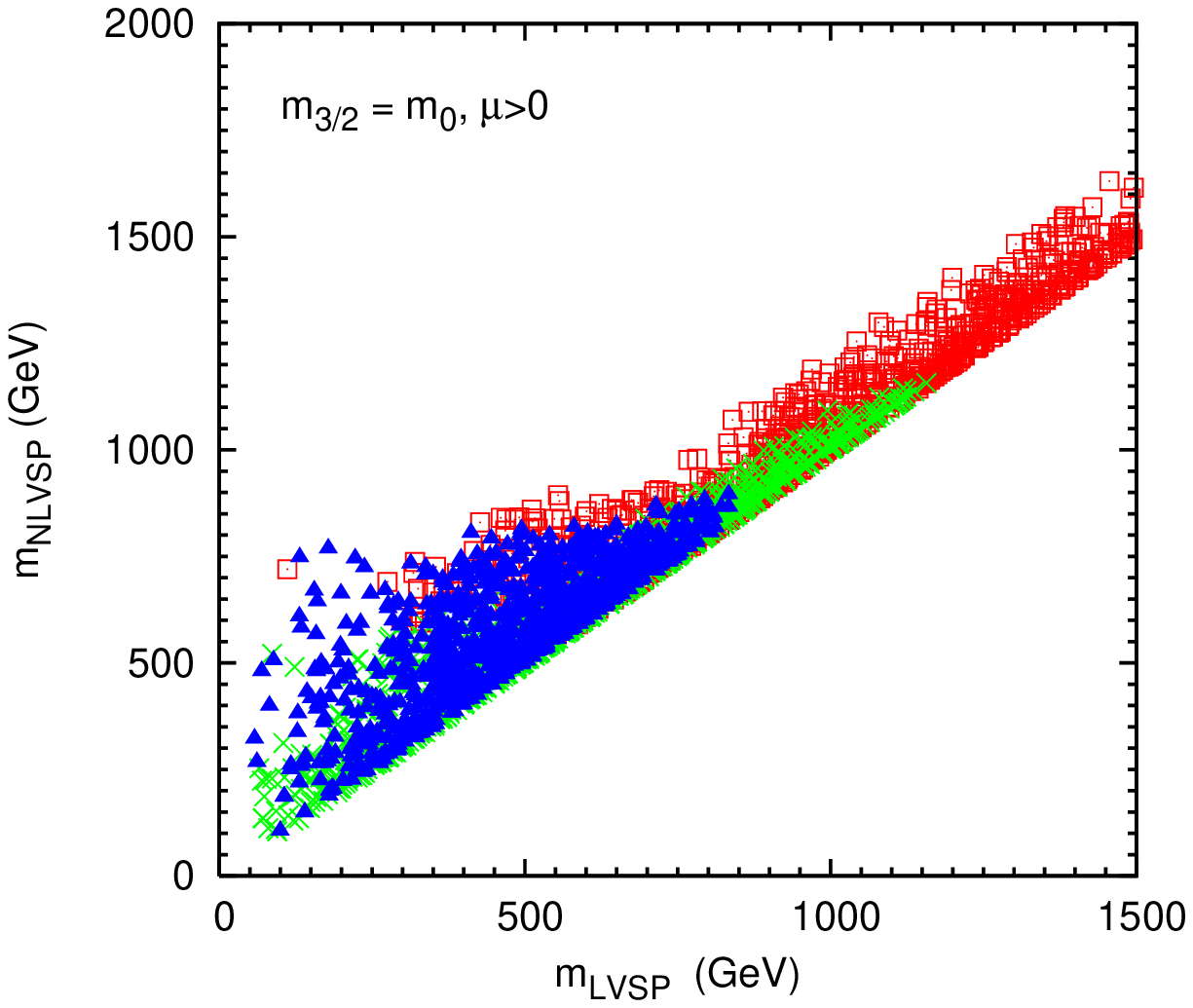}
\end{center}
\caption{Scatter plots of the masses of the lightest visible 
supersymmetric particle (LVSP) and next-to-lightest visible 
supersymmetric particle (NVSP) in (a) the CMSSM, (b) a model with 
non-universal Higgs scalar masses, (c) general scalar masses - all 
assuming a neutralino LSP - and (d) the CMSSM with $m_{3/2} = m_0$, in 
which case the LSP may be the gravitino~\protect\cite{scatter}.} 
\label{fig:scatter} 
\end{figure}

Our present theoretical ignorance means that we do not know the scale of
supersymmetry breaking, and therefore does not yet permit us to fix the
scale at which a LC could be certain of observing any supersymmetric
particles. Fig.~\ref{fig:scatter} displays a set of scatter plots of the
masses of the lightest visible and next-to-lightest visible supersymmetric
particles (LVSP and NVSP, respectively) that could be detected directly at
a future LC, {\it if} $E_{CM} > 2 m_{LVSP, NVSP}$~\cite{scatter}. For
comparison, the green points are accessible to the LHC, the blue points
provide a suitable density of cold dark matter, and the yellow points are
those where this dark matter might be detectable directly in scattering
experiments. Arguments about the fine-tuning of the electroweak scale (and
the magnitude of the relic dark matter LSP density) suggest that
sparticles might be more `likely' to appear near the lower ends of one of
the ranges shown, but we cannot be sure how much fine-tuning is too
much~\cite{finetune}.

Panel (a) of Fig.~\ref{fig:scatter} is for the CMSSM, whereas panel (b)
relaxes universality for the masses-squared of the Higgs bosons and panel
(c) allows all the scalar masses-squared to vary, requiring only that they
remain positive when renormalized up to the GUT scale. These panels all
assume the gravitino to be sufficiently heavy that the LSP is the lightest
neutralino, whereas panel (d) fixes $m_{3/2} = m_0$, in which case the LSP
may be the gravitino and the next-to-lightest supersymmetric particle
might be a neutralino or stau in some regions.

The general conclusion is that a sub-Tev LC would cover only portions of
the allowed supersymmetric parameter spaces (though these portions might
be favoured by fine-tuning arguments), whereas a 3-TeV LC would cover most
of the allowed parameter spaces~\cite{scatter}.

Returning to the CMSSM, Fig.~\ref{fig:Omega}(b) also displays the numbers
of different species of supersymmetric particles that could be seen at a
3- or 5-TeV LC. Such a machine would stand a chance of observing `all' the
sparticles, and would also be able measure in detail the heavier
sparticles -- such as squarks and the heavier gauginos and Higgsinos --
better than the LHC. Studies of heavy slepton production have confirmed
that one could, for example, measure the smuon decay spectrum and infer
the smuon and LSP mass (to $\pm 2.5\%$ and $3\%$, respectively), despite
the greater amount of beamstrahlung inevitable at such a higher-energy
LC~\cite{CLICphys}. With such measurements, one could extend to higher
masses the game of checking GUT and superstring predictions for the
unification of sparticle masses~\cite{BPZ}.

One may safely conclude that any LC above the sparticle threshold would be
very interesting, and that a multi-TeV LC would have considerable added
value in many supersymmetric scenarios, even assuming the prior
construction of a sub-TeV LC.

\section{Extra Dimensions}

If Nature is not wise enough to choose supersymmetry, what alternatives
might she choose? Extra dimensions were first suggested by Kaluza and
Klein in scenarios for the unification of gravity and electromagnetism.
More recently, it has been realized that they are required for the
consistency of string theory, and it was observed that they could help
unify the strong, weak and electromagnetic forces with gravity if at least
one of the extra dimensions is somewhat larger than the Planck
length~\cite{HW}. Larger extra dimensions, around an inverse TeV, could be
the origin of supersymmetry breaking~\cite{Ant}, and even larger extra
dimensions have been postulated in attempts to reformulate the mass
hierarchy problem~\cite{Dim}.

\begin{figure}[htb]
\begin{center}
\epsfig{figure=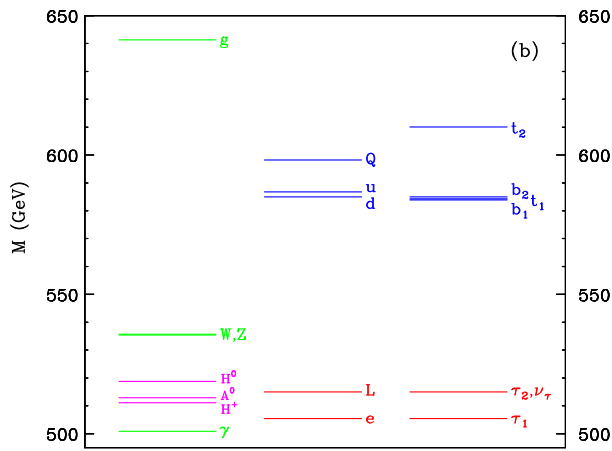,width=8cm}
\end{center}
\caption{The spectrum of Kaluza-Klein recurrences possible an a model 
with universal extra dimensions~\protect\cite{UED}. There are qualitative 
resemblances to a supersymmetric spectrum, whose states have different spins.} 
\label{fig:UED}
\end{figure}

Extra dimensions could be wrapped around in an $S_1$ geometry, as
postulated by Kaluza and Klein (KK), or they could be warped, as
postulated by Randall and Sundrum (RS)~\cite{RS}. Such models may predict
a very rich electron-positron annihilation spectrum due to RS recurrences.
One of the most interesting possibilities is that there are universal
extra dimensions~\cite{UED}. In this case, as seen in Fig.~\ref{fig:UED},
the KK spectrum would look disconcertingly similar to a supersymmetric
spectrum, except that the spins of its KK recurrences would differ from
those of supersymmetric partners of the Standard Model particles. A
sufficiently energetic LC would be ideally placed to measure their spins,
which would be more challenging at the LHC. In such a scenario with
universal extra dimensions, the lightest KK particle would be stable, and
a possible candidate for the astrophysical dark matter~\cite{UEDdm}.

\section{Summary}

As discussed above, there are (still) many good reasons to expect new
physics in the TeV range. However, we shall not know what form this new
physics takes, and what is its energy scale, before the LHC starts
providing results. As emphasized repeatedly, LCs above thresholds for new
physics will provide tremendous added value. At least until we know
(presumably from the LHC) where the threshold(s) for new physics may be,
it is surely advisable to maintain flexibility in the maximum energy which
such LCs could reach.

\end{document}